\documentclass[pra,aps,nofootinbib, print,showpacs,showkeys]{revtex4-2}
\usepackage{amsmath,amssymb,amsfonts,color,graphicx,graphics,latexsym,placeins,epsfig}
\usepackage{footmisc}
\usepackage[compatibility=false]{caption}
\usepackage{subcaption}
\usepackage{bbold}
\usepackage{mathtools}
\usepackage{enumerate}
\usepackage{soul}
\usepackage{hyperref}
\usepackage{tasks}
\usepackage{slashed}
\usepackage{amsthm,bm}
\usepackage{multirow}

\DeclareMathOperator{\sech}{sech}
\usepackage{natbib}
\usepackage{float}

\usepackage[absolute]{textpos}
\begin{document}

\title{\Large \bf Shadow of higher dimensional collapsing dark star and blackhole}

\author{Sagnik Roy} \email{roysagnik134@gmail.com}
\author{Soham Chatterjee} \email{avich1010@gmail.com}
\author{Ratna Koley} \email{ratna.physics@presiuniv.ac.in}
\affiliation{Department  of  Physics,  Presidency  University,  86/1 College Street, Kolkata  700073,  India}

\begin{abstract}
    The shadow of a black hole or a collapsing star is of great importance as we can extract important properties of the object and of the surrounding spacetime from the shadow profile. It can also be used to distinguish different types of black holes and ultra compact objects. In this work, we have analytically calculated the shadow of a higher dimensional collapsing dark star, described by higher dimensional Vaidya metric, by choosing a slightly generalized version of Misner–-Sharp mass function. 
    We have also numerically investigated the properties of the shadows of the black holes and the collapsing stars for a slightly more general mass function. Examining the potential influence of extra spatial dimensions on the shadow, we have explored the possibility of distinguishing higher dimensions from the standard four-dimensional spacetime.
\end{abstract}


\maketitle
\newpage
\section{Introduction}

\emph{Photon sphere} plays a very important role in the study of light trajectories around a blackhole. Light passing near a black hole gets deflected by the strong gravitational field in such a way that it moves along a circular orbit around the black hole and for the case of a spherically symmetric black hole, these circular light trajectories form a sphere called the \emph{photon sphere}. Observation of a black hole against the backdrop of light sources reveals a black disc -- known as the \emph{shadow} of a black hole. For a Schwarzschild black hole, the shadow is circular due to the spherical symmetry of the space time and the light rays spiral in a orbit with radius $r=3m^{bh}$, where $m^{bh}$ is the mass parameter of the black hole. On the other hand, for a rotating Kerr black hole the shadow gets deformed and it is flattened on one side. 
In the minimal set up, an observer will see the shadow if the space around the black hole is filled with light rays and there lies no light source between the observer and the black hole. All past-oriented light rays that starts at observer's position can be categorised into two classes: (i) light rays getting deflected by the black hole and meeting with a source on their way and (ii) light rays going directly to the horizon. We thus assign brightness to the first class of light rays and darkness to the second group. There is a third category of light rays in between these two aforementioned classes and they asymptotically spiral in a photon sphere. \\

In 2000, Falcke et. al. \cite{Falcke_2000} proposed that a shadow can be observed practically and in 2019, Event Horizon Telescope (EHT) \cite{EHT1,EHT2,EHT3,EHT4,EHT5} successfully observed the gravitational light deflection by the supermassive black hole of M87. This is a major breakthrough in this field, followed by the observation of the shadow of supermassive black hole Sagittarius A*  in the center of our Milky Way in 2022 \cite{sg_A_shadow}. Testing of compact gravitational objects, arising out of different theoretical propositions is being done through this new window of observation. \\

Analytical calculations of shadow is generally done by considering a black hole against the backdrop of light sources with the assumption that there is no source between the black hole and the observer. 
The  eternal black holes have been the popular choice for these calculations so far. In 1966, Synge \cite{Synge_Shadow} analytically calculated the angular radius of the shadow of a Schwarzschild black hole and Bardeen \cite{Black_holes} found the shape of the shadow of a Kerr black hole for an observer at infinity by choosing eternal black holes. According to our present knowledge, the black holes are formed by gravitational collapse and evolve by accretion of surrounding matter or mergers with other black holes. Thus we find it quite interesting to explore the shadow of evolving systems. Visual appearance of a star collapsing through its gravitational radius was first studied by Ames and Thorne \cite{Ames_Thorne}, after that Jaffe \cite{other_spectra_papers1}, Lake and Roeder \cite{other_spectra_papers2} and Frolov et. al. \cite{Frolov_2007} studied the frequency shift of light coming from the surface of a collapsing non-transparent star. On the other hand, work by Kong et. al. \cite{Kong_2014, Kong_2015} and Ortiz et. al. \cite{Ortiz_2015, ortiz2015observational} dealt with the frequency shift of light passing through a collapsing transparent star, thereby contrasting the collapse to a black hole with the collapse to a naked singularity. General properties of photon sphere of a static, spherically symmetric spacetime is discussed in \cite{Claudel_2001}. A detailed review of analytical calculations of black shadow is presented in \cite{Perlick_review}. Recently Schneider and Perlick \cite{shadow_dark_star} studied the shadow of a dark and non-transparent collapsing star. Also Solanki and Perlick \cite{4D_Vaidya_shadow} studied the shadow of a time-dependent black hole described by Vaidya metric. The above-mentioned works are all in $4$-dimensions. However, the shadows of different kinds of black holes in higher dimensions have also been studied with great interest by several physicists. Singh and Ghosh obtained the shadow of a Schwarzschild--Tangherlini black hole \cite{Schwarzschild_Tangherlini_shadow}, while Papnoi et. al. \cite{Myers_Perry_5D} studied the shadow of $5$-dimensional rotating Myers--Perry black hole and its \textit{regular} version respectively. The shadows of other types of higher dimensional black holes and the effect of extra dimensions on the shadow have been explored in \cite{Higher_dim_effect_on_BH_shadow,Higher_dim_BH_shadow1,Higher_dim_BH_shadow2,Higher_dim_BH_shadow3, KK_shadow}.\\

In this work, we consider a \emph{higher dimensional collapsing dark star} emitting radiation and non-relativistic matter particles. In the situation when the amount of radiation is too feeble to be detected, the star will appear as a non-transparent, black disc against the background of light sources. In order to explore the evolution of the black disc, we choose the generalized Vaidya metric which is a non-vacuum solution of the Einstein's field equation around a spherically symmetric body and generalized it to the higher dimensions. The photon sphere and shadow profile have been studied thereafter. In a recent work by Banerjee et. al. \cite{sg_A_shadow_higherdim} it has been argued that the observed shadow of the M87 and Sgr A* may carry a signature of extra dimension through a tidal charge as the hair. It was shown that the observed shadow and the image diameter of the Sgr A* always predict a non-zero value of the tidal charge parameter within the allowed range of angular diameter of the shadow. An independent probe of the tidal charge will be required to come to a conclusion about the extra dimension. 
However, a more complete study is due where one should consider the effect of the environment on the shadow profile in the higher dimensional scenario. The degeneracy in the existing explanations of the shadow profiles can only be lifted by means of independent experimental results. 
In this work we have adopted a different way through which we explored how the shadow of a Vaidya collapsing star (as well as of a Vaidya black hole) strongly depends on the mass and the dimensionality of spacetime. We have argued that the position of the photon sphere will remain unchanged by changing the spacetime dimension ($D$) and accordingly the mass function. Therefore, the mass turns out to be an important parameter characterizing the “same” shadow in different dimensions. We have utilized this property to comment on how the mass of a compact object may be utilized to fix the dimensionality of the spacetime. It is worthy mentioning -- this method is useful with the aid of astrophysical observations that will provide us with an independent probe of the mass which in turn 
lead us to the conclusion. The studies done so far in this direction consider only the vacuum spacetime. Our work is based on Vaidya spacetime, which is a more realistic choice to model a black hole as well as a collapsing star. Obtaining the shadow of such a collapsing star both analytically and numerically, we have shown how the mass of the star can be used to speculate a signature of the extra spatial dimensions. \\

The work is organised in the following way. In Section II, we have discussed the generalized $D$-dimensional Vaidya metric in Eddington--Finkelstein like coordinates and the generalized mass function. A new conformal symmetry has been identified through a coordinate transformation. The null geodesic has been studied in this new coordinate for a special type of mass parameter. Section III deals with the angular radius and the shape of a higher dimensional evaporating black hole in outgoing Vaidya metric. We have used a slightly generalized version of Misner--Sharp mass function. The escape angle of a higher dimensional Vaidya black hole has been calculated in the frame of a static observer. In Section IV, we first obtained the shadow of a collapsing star for a static observer in the transformed frame and then reverted back to the original coordinates (Eddington--Finkelstein-like coordinates). Numerical analysis of the photon sphere and the shadow profile is shown in Section V. Next we explored the detection possibilities of extra dimensions using the shadow of a collapsing star. Finally concluded in Section VII.

\section{Null Geodesics in Higher dimensional Vaidya Spacetime}

The matter fields present in the nature can be classified in general into two broad categories -- (i) with energy-momentum tensor having one timelike and three spacelike eigenvectors (this includes dust, perfect fluid, called type-I matter field) and (ii) with energy-momentum tensor having   double null eigenvectors (which includes radiation, null dust, called type-II matter field). The most general spherically symmetric metric for any arbitrary combination of these two types of matter fields in outgoing Eddington--Finkelstein-like coordinates is given by \cite{Ojako_ref-28,Misner_Sharp_mass_function},
\begin{equation}
    ds^2 = -e^{2\psi_{4}(u,r)}\left[1-\dfrac{2m_{4}(u,r)}{r}\right] du^2 - 2e^{\psi_{4}(u,r)} dudr + r^2 d\Omega_{2}^2
\end{equation}
where $m_{4}(u,r)$ is the Misner--Sharp mass function in $4$-dimensions \cite{Misner_Sharp_mass_function,Ojako_etal_conformal_sym_Vaidya}. This tells about the amount of energy inside radial distance $r$ at retarded time $u$ and $d\Omega_2^2$ is the metric on a $2$-dimensional unit sphere. Wang and Wu \cite{Wang_Wu} proved that a specific combination of the  two types of matter fields mentioned above can give rise to a generalized Vaidya metric as follows
\begin{equation}
    ds_4^2 = -\left(1-\dfrac{2m_4(u,r)}{r}\right)du^2 - 2 dudr + r^2 d\Omega_{2}^2
\end{equation}
This is a straight forward generalisation of the original Vaidya metric \cite{Vaidya_original_paper}. The same metric was generalized for $D$-dimensions by Iyer and Vishveshwara in \cite{VaidyaMetric5D} (further generalisation for charged blackholes can be found in \cite{VaidyaMetric5D_with_charge}),
\begin{equation}
    ds_D^2 = -\left(1-\dfrac{2m_D(u,r)}{r^{D-3}}\right)du^2 - 2 dudr + r^2 d\Omega_{D-2}^2 \label{metric}
\end{equation}
where $d\Omega_{D-2}^2$ denotes the metric on a $(D-2)$-dimensional unit sphere, defined by,
\begin{equation}
    d\Omega_{D-2}^2 = d\theta_1^2 + \displaystyle\sum_{i=2}^{D-2} \left(\displaystyle\prod_{j=1}^{i-1} \sin{^2\theta_j}\right) d\theta_i^2
\end{equation}
and $m_D(u,r)$ is the $D$-dimensional Misner--Sharp mass function \cite{Schwarzschild_Tangherlini_shadow} defined as
\begin{equation}
    m_D(u,r) = \dfrac{4\pi^{-(D-3)/2}}{(D-2)} \Gamma\left(\dfrac{D-1}{2}\right) G_D M_D(u,r) ~ \propto ~ G_D M_D(u,r)
\end{equation}
where $G_D$ is Newton's gravitational constant in $D$-dimensions and $M_D(u,r)$ is the mass of the star for its radius $r$ at time $u$. Let us consider that the star is collapsing by radiating shell of null dusts and other matter particles. We take  \emph{outgoing} Eddington--Finkelstein time in the metric. Also throughout the paper, we consider a \emph{soft} version of the Oppenheimer--Snyder model \cite{OppenheimerSnyder}, i.e. a very small amount of outward pressure resides at the surface of the dark star, so that a radial timelike inward geodesic of a free particle can represent the collapse of the star surface. A careful examination reveals there is no other conformal Killing vector field except $\frac{\partial}{\partial \phi}$, where we have set $\theta_{D-2} = \phi$. Therefore, it is \emph{almost impossible} (barring some exceptions) to do any analytical calculation of the shadow profile due to lack of enough constants of motion. In that case, one has to import some kind of numerical techniques (which we will do later). 

In the following we will show that analytically one can proceed with a special form for $m_D(u,r)$. Ojako et. al. stated in \cite{Ojako_etal_conformal_sym_Vaidya} that there exists a homothetic Killing vector field, 
for which the spacetime is self-similar and the mass function for the same is as follows. 
\begin{equation}
    m_4(u,r) = \displaystyle\sum_{n\in\mathbb{Z}} \dfrac{a_n u^n}{r^{n-1}}
\end{equation}

where $\mathbb{Z}$ represents the set of all integers on the real line. In the very similar fashion the mass function will take the following form in $D$-dimensions
\begin{equation}
    m_D(u,r) = \displaystyle\sum_{n\in\mathbb{Z}} \dfrac{a_n u^n}{r^{n+3-D}} \label{m_D1}
\end{equation}

Obviously, the values of $a_n$'s should be so chosen that the convergence of the series is confirmed. It is trivial to notice  if all $a_n$'s are zero except for $a_1$, the mass function will correspond to constant mass accretion or expel in $4$-dimensions. We now perform a coordinate transformation to non--angular coordinates, $(u,r) \longrightarrow (T,R)$, following \cite{4D_Vaidya_shadow}, to identify a new conformal symmetry,
\begin{equation}
    u = r_0 ~ e^{T/r_0} ~~\text{and}~~ r = R ~ e^{T/r_0} \label{ur_to_TR}
\end{equation}
where $r_0$ is an arbitrary constant with the dimension of length. In the transformed coordinates, the line element \eqref{metric} becomes,
\begin{equation}
    ds_D^2 = e^{2T/r_0} \left\{-\left(1 - \displaystyle\sum_{n\in\mathbb{Z}} \dfrac{a_n r_0^n}{R^n} - \dfrac{2R}{r_0}\right)dT^2 - 2 dTdR + R^2 d\Omega_{D-2}^2\right\} \label{metric_new}
\end{equation}
The Lagrangian density corresponding to a massless particle (on the equatorial plane i.e. $\theta_i=\pi/2$ for $1 \le i \le D-3$) in this set of new coordinates is given by
\begin{equation}
    \mathcal{L}_D = e^{2T/r_0} \left\{-\left(1 - \displaystyle\sum_{n\in\mathbb{Z}} \dfrac{a_n r_0^n}{R^n} + \dfrac{2R}{r_0}\right)\Dot{T}^2 - 2 \Dot{T}\Dot{R} + R^2 \Dot{\phi}^2 \right\} = 0 \label{new_Lagrangian1}
\end{equation}

It is trivial to identify the conformally conserved angular momentum from the above equation. 
\eqref{new_Lagrangian1}:
\begin{equation}
    L = R^2 e^{2T/r_0} \Dot{\phi} \label{conserveL}
\end{equation}
In addition to this we also get $\frac{\partial}{\partial T}$ from Eq. \eqref{new_Lagrangian1}, as a conformal Killing vector field leading to a constant of motion $\mathcal{E}$, along every null geodesic. 
\begin{equation}
    \mathcal{E} = e^{2T/r_0} \left\{-\left(1-\displaystyle\sum_{n\in\mathbb{Z}} \dfrac{a_n r_0^n}{R^n} - \dfrac{2R}{r_0}\right)\Dot{T} - \Dot{R}\right\} \label{conserveE}
\end{equation}
It is also apparent from Eq. \eqref{new_Lagrangian1} that for radial light rays $\left(\Dot{\phi}=0\right)$ one can have following 
relations for $T$.
\begin{equation}
    T = T_0 ~~\text{(constant) ~ or}~~ T = T_0 - 2 \displaystyle\int \left(1 - \displaystyle\sum_{n\in\mathbb{Z}} \dfrac{a_n r_0^n}{R^n} - \dfrac{2R}{r_0}\right)^{-1} dR
\end{equation}
where the first equation is valid for outgoing and the second stands  for incoming radial light rays.
We now explore the geodesic motion of massless particles in this background.  For this purpose, 
we have to solve two coupled equations, one appearing from the Lagrangian density in Eq. \eqref{new_Lagrangian1} and the other from the combined version of the conservation laws in Eqs. \eqref{conserveL} and \eqref{conserveE}:
\begin{eqnarray}
    \left(1 - \displaystyle\sum_{n\in\mathbb{Z}} \dfrac{a_n r_0^n}{R^n} - \dfrac{2R}{r_0}\right) \left(\dfrac{dT}{d\phi}\right)^2 + 2~\dfrac{dT}{d\phi} \dfrac{dR}{d\phi} = R^2 \label{dTdphi_dRdphi_1} \\
   \left(1-\displaystyle\sum_{n\in\mathbb{Z}} \dfrac{a_n r_0^n}{R^n} - \dfrac{2R}{r_0}\right) \dfrac{dT}{d\phi} + \dfrac{dR}{d\phi} = \dfrac{\mathcal{E} R^2}{L} \label{dTdphi_dRdphi_2}
\end{eqnarray}
Solution of the above set of equations \eqref{dTdphi_dRdphi_1} and \eqref{dTdphi_dRdphi_2}, will provide us with the motion of a test photon on the equatorial plane in this spacetime. Simplifying the above equations we obtain 
\begin{eqnarray}
    \dfrac{dT}{d\phi} &=& \dfrac{-\dfrac{\mathcal{E}R^2}{L} \mp \sqrt{\dfrac{\mathcal{E}^2R^4}{L^2} + \displaystyle\sum_{n\in\mathbb{Z}} \dfrac{a_n r_0^n}{R^{n-2}} + \dfrac{2R^3}{r_0} - R^2}}{1 - \displaystyle\sum_{n\in\mathbb{Z}} \dfrac{a_n r_0^n}{R^n} - \dfrac{2R}{r_0}} \label{dTdphi} \\
   \dfrac{dR}{d\phi} &=& \pm \sqrt{\dfrac{\mathcal{E}^2R^4}{L^2} + \displaystyle\sum_{n\in\mathbb{Z}} \dfrac{a_n r_0^n}{R^{n-2}} + \dfrac{2R^3}{r_0} - R^2} \label{dRdphi}
\end{eqnarray}
Note that the above equations will give $T$ and $R$ as function of $\phi$. We do not find the geodesics explicitly. The expression in Eq. \eqref{dRdphi} has been used in Section \ref{Shadow_BH} to calculate the shadow of an evaporating black hole  as well as the shadow of a collapsing dark star in Section \ref{Collapsing_star_shadow_(T,R)}.

\section{Shadow Profile of a Higher-dimensional Black Hole described by Vaidya Metric} \label{Shadow_BH}

Let us consider a higher dimensional evaporating black hole. In four dimensions an evaporating black hole can be -- very idealistically -- modelled by a $4D$ Vaidya metric with decreasing mass function \cite{evaporating-1_Hiscock-1,evaporating-2_Hiscock-2, evaporating-3_Kuroda, evaporating-4_Beciu}. We assume this model remains  valid in higher dimensions also. The shadow profile of a $4$D Vaidya spacetime was studied by Solanki and Perlick \cite{4D_Vaidya_shadow} for a linearly increasing and decreasing Misner--Sharp mass function.  They have shown for a black hole with linearly increasing mass there exist two horizons. This is in contrast to the linearly decreasing mass, which admits only one horizon. The angular radius of an evaporating Vaidya black hole shadow is shown to be time-independent for a conformally static observer though the area of the photon sphere is decreasing. On the contrary to this work, we have generalized the functional dependence of the mass function slightly as given in Eq. \eqref{m_D1}. \\

In this section, we find the expression for the photon sphere and angular radius of a Vaidya black hole as seen by a static observer in $(T,R)$ frame. Let us identify the tetrad basis for the metric in Eq. \eqref{metric},
\begin{eqnarray}\label{tetrad}
    e_T &=& e^{-T/r_0} \left(1-\displaystyle\sum_{n\in\mathbb{Z}} \dfrac{a_n r_0^n}{R^n}-\dfrac{2R}{r_0}\right)^{-1/2} \dfrac{\partial}{\partial T} \\
    e_R &=& e^{-T/r_0} \left[ \left(1-\displaystyle\sum_{n\in\mathbb{Z}} \dfrac{a_n r_0^n}{R^n}-\dfrac{2R}{r_0}\right)^{-1/2} \dfrac{\partial}{\partial T} ~-~ \left(1-\displaystyle\sum_{n\in\mathbb{Z}} \dfrac{a_n r_0^n}{R^n}-\dfrac{2R}{r_0}\right)^{1/2} \dfrac{\partial}{\partial R} \right] \\
    e_\phi &=& \dfrac{e^{-T/r_0}}{R} \dfrac{\partial}{\partial \phi} 
  \end{eqnarray}
  
We have also set $\theta_{D-2} = \phi$ as the azimuthal angle of the coordinate system and have restricted ourselves on the equatorial plane (i.e. $\theta_i=\pi/2$ for $1 \le i \le D-3$) of the collapsing star.\\

We now consider a null geodesic $(T(s),R(s),\phi(s))$ on the equatorial plane, where $s$ is the affine parameter along the geodesic of a massless particle.  Let us expand the tangent vector with respect to the tetrad \eqref{tetrad}. Since the tangent vector is light-like, the expansion can be written in terms of the angle $(\alpha)$ between the photon geodesic and the radial direction in the rest system of the observer in the following manner.
\begin{equation}
    \Dot{T}~\dfrac{\partial}{\partial T} + \Dot{R}~\dfrac{\partial}{\partial R} + \Dot{\phi}~\dfrac{\partial}{\partial \phi} = \kappa \left(e_T + e_R\cos{\alpha} + e_\phi\sin{\alpha}\right) \label{null_tangent}
\end{equation}
where $\kappa$ is a scale factor $(\kappa>0)$, $\alpha$ denotes the celestial coordinate and $\Dot{T}$ represents $\frac{dT}{ds}$ and so on. Comparison of the coefficients of $\frac{\partial}{\partial T}$, $\frac{\partial}{\partial R}$ and $\frac{\partial}{\partial \phi}$ on both sides of Eq. \eqref{null_tangent} results in
\begin{eqnarray}\label{Tdot_Rdot_phidot}
    \Dot{T} &=& \phantom{+} \kappa ~ e^{-T/r_0} \left(1-\displaystyle\sum_{n\in\mathbb{Z}} \dfrac{a_n r_0^n}{R^n}-\dfrac{2R}{r_0}\right)^{-1/2} \left(1+\cos{\alpha}\right) ~ , \\
    \Dot{R} &=& -\kappa ~ e^{-T/r_0} \left(1-\displaystyle\sum_{n\in\mathbb{Z}} \dfrac{a_n r_0^n}{R^n}-\dfrac{2R}{r_0}\right)^{1/2} \cos{\alpha} ~~ \text{and} \\
    \Dot{\phi} &=& \phantom{+} \dfrac{\kappa ~ e^{-T/r_0}}{R} ~ \sin{\alpha}
\end{eqnarray}
From the last two equations above we can write 
\begin{equation}
    \left(\dfrac{dR}{d\phi}\right)^2 =  R^2\cot{^2\alpha} \left(1-\displaystyle\sum_{n\in\mathbb{Z}} \dfrac{a_n r_0^n}{R^n}-\dfrac{2R}{r_0}\right) \label{dRdphi_squared}
\end{equation}
Further comparing Eqs. \eqref{dRdphi} and \eqref{dRdphi_squared} we obtain
\begin{equation}
    \sin{^2\alpha} = \dfrac{L^2}{\mathcal{E}^2R^2} \left(1-\displaystyle\sum_{n\in\mathbb{Z}} \dfrac{a_n r_0^n}{R^n}-\dfrac{2R}{r_0}\right) \label{sin2_alpha_BH}
\end{equation}

where $\mathcal{E}/L$ represents some proportionality of the impact parameter of a photon within the black hole spacetime. Here it is a constant of motion. However, in practice it is not that much reliable while observing a shadow. Instead a photon sphere or something like that may be a more useful parameter. That is why we express $L^2/\mathcal{E}^2$ in terms of the position of the photon sphere (or any equivalent quantity) of this spacetime. Evaluating the condition $\frac{dR}{d\phi}=0$ one can get the extremum points of light paths, $R = R_m$ \cite{4D_Vaidya_shadow} from the following equation 
\begin{equation}
    \dfrac{L^2}{\mathcal{E}^2} = R_m^2 \left(1 - \sum_{n\in\mathbb{Z}} \dfrac{a_n r_0^n}{R_m^n} - \dfrac{2R_m}{r_0}\right)^{-1} \label{L^2/E^2}
\end{equation}
The radial coordinate $R_m$ will be the position of photon sphere if it satisfies the condition $\left.\frac{d^2R}{d\phi^2}\right|_{R=R_m} = 0$ \cite{4D_Vaidya_shadow}. Imposing this we get
\begin{equation}
    \displaystyle\sum_{n\in\mathbb{Z}} \dfrac{a_n (\frac{n}{2}+1) r_0^n}{R_m^n} + \dfrac{R_m}{r_0} = 1
\end{equation}
The solution of this equation will give the position of photon sphere. Now by using Eqs. \eqref{sin2_alpha_BH} and \eqref{L^2/E^2}, we have
\begin{equation}
    \sin{\alpha} = \dfrac{R_m}{R} \sqrt{\dfrac{1-\sum \frac{a_n r_0^n}{R^n}-\frac{2R}{r_0}}{1 - \sum \frac{a_n r_0^n}{R_m^n} - \frac{2R_m}{r_0}}}
\end{equation}
An observer situated at $R = R_O$ will measure the angular radius of the shadow of a $D$-dimensional evaporating Vaidya black hole  as 
\begin{equation}
    \sin{\alpha_{sh}^{bh}} = \dfrac{R_m}{R_O} \sqrt{\dfrac{1-\sum \frac{a_n r_0^n}{R_O^n}-\frac{2R_O}{r_0}}{1 - \sum \frac{a_n r_0^n}{R_m^n} - \frac{2R_m}{r_0}}} \label{sin alpha_sh}
\end{equation}

where the subscript $sh$ and the superscript $bh$ in $\alpha$ will mean the shadow of the black hole for the corresponding observer. This formula clearly depicts that the angular radius of this shadow is time independent for an observer on the constant--$R$ line \textit{i.e.} the boundary of the shadow profile, which will be circular in shape due to the spherical symmetry of the spacetime, will not evolve over time $T$ and this is quite obvious from the fact that the spacetime is conformally static  in $(T,R)$-frame, as shown earlier.

\section{Shadow of a Collapsing dark Star for a Static Observer}\label{SectionIV}

We start by considering the free fall of a radially ingoing massive particle (residing just above the star surface)
on the equatorial plane of this spacetime. The star is collapsing in a spherically symmetric way, so if we consider infinitely many particles, all at the same distance from the origin of the coordinate system, then all of them will follow the similar infalling geodesics, which will in turn represent the infalling trajectory of the star surface. Thus to mimic the collapse of the star surface, it is enough to consider a timelike geodesic depicting free infall towards the origin \cite{shadow_dark_star}. 
 The Lagrangian and the conformally conserved energy of such a vertically infalling massive test particle on the equatorial plane of this spacetime are given by Eqs. \eqref{new_Lagrangian1} and \eqref{conserveE} for $\Dot{\phi}=0$. 
The Lagrangian of a vertically infalling massive particle following Eq. \eqref{new_Lagrangian1} implies
\begin{equation}
    \left(1 - \displaystyle\sum_{n\in\mathbb{Z}} \dfrac{a_n r_0^n}{R^n} - \dfrac{2R}{r_0}\right)\Dot{T}^2 + 2 \Dot{T}\Dot{R} = e^{-2T/r_0} \label{Lagrangian_infall}
\end{equation}
and the conformal constant of motion (from Eq. \eqref{conserveE}) corresponding to the Killing vector $\frac{\partial}{\partial T}$ takes the form 
\begin{equation}
    -\left(1-\displaystyle\sum_{n\in\mathbb{Z}} \dfrac{a_n r_0^n}{R^n} - \dfrac{2R}{r_0}\right)\Dot{T} - \Dot{R} = e^{-2T/r_0} \mathcal{E}. \label{conserveE_infall}
\end{equation}

In the following, we restrict ourselves to the condition $\Dot{u} > 0$, which depicts that time is future oriented and $\Dot{r} < 0$, {\textit{i.e.}} the particle is infalling. Recalling the inverse coordinate transformations: $T = r_0 \ln{\dfrac{u}{r_0}} ~~\text{and}~~ R = r_0\dfrac{r}{u} $ we get 
\begin{eqnarray}
    %
    \Dot{T} &=& r_0 \dfrac{\Dot{u}}{u}  ~~\text{and}~~ \Dot{R} = \dfrac{r_0}{u^2} \left(u\Dot{r}-r\Dot{u}\right).
\end{eqnarray}

The conditions of infall of a massive particle mentioned above automatically imply $\Dot{T} > 0$ and $\Dot{R} < 0$. One does not find any solution set for $\Dot{T}$ and $\Dot{R}$ admitting this infalling condition while solving Eqs. \eqref{Lagrangian_infall} and \eqref{conserveE_infall}. However, there is no physical reason behind the non-existence of an infalling timelike geodesic in this spacetime as we know Vaidya spacetime can describe the infall of massive particles. The spacetime under consideration is just a coordinate-transformed version of the Vaidya metric. Therefore, it should possess an infalling timelike geodesic which can represent the radial infall of the star surface and hence star collapse. One good way is to identify the conformally conserved energy, $E$ as $-\mathcal{E} = -\partial \mathcal{L}_D/\partial \Dot{T}$. Accordingly, we get
\begin{equation}
    \left(1-\displaystyle\sum_{n\in\mathbb{Z}} \dfrac{a_n r_0^n}{R^n} - \dfrac{2R}{r_0}\right)\Dot{T} + \Dot{R} = e^{-2T/r_0} E \label{conserve-E_infall}
\end{equation}
Solving Eq. \eqref{Lagrangian_infall} and Eq. \eqref{conserve-E_infall}, we obtain (for  $\Dot{T} > 0$ and $\Dot{R} < 0$)
\begin{eqnarray}
    \Dot{T} &=& e^{-2T/r_0} \left\{E + \sqrt{E^2 - e^{2T/r_0} \left(1-\displaystyle\sum_{n\in\mathbb{Z}} \dfrac{a_n r_0^n}{R^n} - \dfrac{2R}{r_0}\right)}\right\}\left(1-\displaystyle\sum_{n\in\mathbb{Z}} \dfrac{a_n r_0^n}{R^n} - \dfrac{2R}{r_0}\right)^{-1} \label{dTdtau} \\
    \Dot{R} &=& -e^{-2T/r_0} \sqrt{E^2 - e^{2T/r_0} \left(1-\displaystyle\sum_{n\in\mathbb{Z}} \dfrac{a_n r_0^n}{R^n} - \dfrac{2R}{r_0}\right)} \label{dRdtau}
\end{eqnarray}

Further simplification of Eqs. \eqref{dTdtau} and \eqref{dRdtau} yields
\begin{equation}
    \dfrac{dR}{dT} = e^{-2T/r_0} E^2 - \left(1-\displaystyle\sum_{n\in\mathbb{Z}} \dfrac{a_n r_0^n}{R^n} - \dfrac{2R}{r_0}\right) - e^{-2T/r_0}E^2 \sqrt{1 - \dfrac{e^{2T/r_0}}{E^2} \left(1-\displaystyle\sum_{n\in\mathbb{Z}} \dfrac{a_n r_0^n}{R^n} - \dfrac{2R}{r_0}\right)}
\end{equation}

The relative velocity between a radially infalling observer and a static observer outside the star in $(T,R)$ frame is found to be \footnote{This has been discussed in \cite{infalling_velocity} for a general static $4$-dimensional case and the formula for the relative velocity $\left(\Vec{v} \equiv \left(v_R,0,0,...\right)\right)$ between a radially infalling observer and a static observer has been derived in \cite{Landau_Lifshitz}; and it turns out that the proof is easily extendable to any higher dimensional static case.}
\begin{equation}
    v_R = \sqrt{1 - \dfrac{e^{2T/r_0}}{E^2} \left(1 - \displaystyle\sum_{n\in\mathbb{Z}} \dfrac{a_n r_0^n}{R^n} - \dfrac{2R}{r_0}\right)}  \label{infall_velocity_our_case}
\end{equation}

We now calculate the orthogonal tetrads for a radially infalling observer in the $(T,R)$ frame using the above results. 
For this purpose it is necessary to generalize the family of orthogonal tetrads (denoted by $\Tilde{e}_\mu$) for any arbitrary observer in $D$-dimensions, moving with velocity $\Vec{v} \equiv \left(v_1,v_2,v_3,...\right)$ with respect to a static frame\footnote{The method of generating the family of orthogonal tetrads in $4$-dimensions with respect to a static frame has been discussed in \cite{infalling_tetrad_paper,infalling_tetrad_book}.}. Careful observations reveal that the extra dimensional coordinates being angular in nature, the number of non-zero components of basis tetrads reduces due to spherical symmetry and we are left with only $T$, $R$ and $\phi$ components. Here the $\phi$ component is identical with its counterpart in $4$-dimensions for obvious reasons.


The orthogonal tetrads for a freely infalling observer with respect to a static frame thus become 
\begin{eqnarray}\label{infall_tetrad}
    \Tilde{e}_T &=& \dfrac{e_T + v_R ~ e_R}{\sqrt{1-v_R^2}} = e^{-2T/r_0} \left[\dfrac{E - \sqrt{E^2-m_*e^{2T/r_0}}}{m_*} \dfrac{\partial}{\partial T} - \sqrt{E^2 - m_*e^{2T/r_0}}~\dfrac{\partial}{\partial R}\right] \\
    \Tilde{e}_R &=& \dfrac{v_R ~ e_T + e_R}{\sqrt{1-v_R^2}} = e^{-2T/r_0} \left[\dfrac{E - \sqrt{E^2-m_*e^{2T/r_0}}}{m_*} \dfrac{\partial}{\partial T} - E~\dfrac{\partial}{\partial R}\right] \\
    \Tilde{e}_\phi &=& e_\phi = \dfrac{e^{-T/r_0}}{R} \dfrac{\partial}{\partial \phi} ~~\text{(on the equatorial plane)}
\end{eqnarray}
where $m_* = 1 - \displaystyle\sum_{n\in\mathbb{Z}} \dfrac{a_n r_0^n}{R^n} - \dfrac{2R}{r_0}$ has been defined for the sake of brevity.

\subsection{Shadow profile of a collapsing star from a static frame in {(T,R)} coordinates}\label{Collapsing_star_shadow_(T,R)}

In order to determine the escape angle for the collapsing star, we will follow exactly the same method, as we did earlier in the case of a static observer (in $(T,R)$ frame) in section \eqref{Shadow_BH}. Let us denote this angle (celestial angle) between the photon geodesic and the radial direction in the rest frame  of the observer by $\Tilde{\alpha}$ and therefore (the scale factor in this case is $\Tilde{\kappa}>0$), 
\begin{equation}
    \Dot{T}~\dfrac{\partial}{\partial T} + \Dot{R}~\dfrac{\partial}{\partial R} + \Dot{\phi}~\dfrac{\partial}{\partial \phi} = \Tilde{\kappa} \left(\Tilde{e}_T + \Tilde{e}_R\cos{\Tilde{\alpha}} + \Tilde{e}_\phi\sin{\Tilde{\alpha}}\right) \label{null_tangent_infall}
\end{equation}
Comparing the coefficients of $\dfrac{\partial}{\partial T}$, $\dfrac{\partial}{\partial R}$ and $\dfrac{\partial}{\partial \phi}$ on both sides of Eq. \eqref{null_tangent_infall}, we have
\begin{equation}
    \left(\dfrac{dR}{d\phi}\right)^2 =  \dfrac{R^2 e^{-2T/r_0}}{\sin{^2\Tilde{\alpha}}} \left(\sqrt{E^2 - m_*e^{2T/r_0}} - E~\cos{\Tilde{\alpha}}\right)^2 \label{dRdphi_squared_infall}
\end{equation}
Now using the expression of $\dfrac{dR}{d\phi}$ from Eq. \eqref{dRdphi}, we have,
\begin{eqnarray}
    \dfrac{E^2R^2}{L^2} - \left(1 - \displaystyle\sum_{n\in\mathbb{Z}} \dfrac{a_n r_0^n}{R^n} - \dfrac{2R}{r_0}\right) = \dfrac{e^{-2T/r_0}}{\sin{^2\Tilde{\alpha}}}& \nonumber \\
    \left\{\sqrt{E^2 - \left(1 - \displaystyle\sum_{n\in\mathbb{Z}} \dfrac{a_n r_0^n}{R^n} - \dfrac{2R}{r_0}\right) e^{2T/r_0}} + E~\cos{\Tilde{\alpha}}\right\}^2& \label{sin2_alpha_BH_infall}
\end{eqnarray}
and finally by substituting $E^2/L^2 = \mathcal{E}^2/L^2$ from Eq. \eqref{L^2/E^2}, we now get,
\begin{eqnarray}
    &\dfrac{R^2}{R_m^2}\left(1 - \displaystyle\sum_{n\in\mathbb{Z}} \dfrac{a_n r_0^n}{R_m^n} - \dfrac{2R_m}{r_0}\right)-\left(1-\displaystyle\sum_{n\in\mathbb{Z}} \dfrac{a_n r_0^n}{R^n} - \dfrac{2R}{r_0}\right) \nonumber \\ 
    &= \dfrac{e^{-2T/r_0}}{\sin{^2\Tilde{\alpha}}} \left\{\sqrt{E^2 - \left(1 - \displaystyle\sum_{n\in\mathbb{Z}} \dfrac{a_n r_0^n}{R^n} - \dfrac{2R}{r_0}\right) e^{2T/r_0}} + E~\cos{\Tilde{\alpha}}\right\}^2
\end{eqnarray}

One can obtain the expression for the celestial angle $\Tilde{\alpha}$ from the above equation. However, that will not be useful in calculating the shadow of a collapsing star for a static observer. So, we skip that part. The shadow is obtained by  considering the null geodesics that are grazing the surface of the star. One can (in principle) determine $R_m$ by equating $R = R_s$ (where $R_s$ is the radius of the star) and $\Tilde{\alpha} = \pi/2$, if such a null geodesic passes through a minimum radius value $R_m$. An observer on the star surface will see its entire sky to be covered by the shadow. Therefore we get from the above equation (45)
\begin{eqnarray}
    \dfrac{R_s^2}{R_m^2} \left(1-\sum_{n\in\mathbb{Z}} \dfrac{a_n r_0^n}{R_m^n} - \dfrac{2R_m}{r_0}\right) - \left(1 - \displaystyle\sum_{n\in\mathbb{Z}} \dfrac{a_n r_0^n}{R_s^n} - \dfrac{2R_s}{r_0}\right) \nonumber \\
    = e^{-2T/r_0} \left\{E^2 - \left(1 - \displaystyle\sum_{n\in\mathbb{Z}} \dfrac{a_n r_0^n}{R_s^n} - \dfrac{2R_s}{r_0}\right) e^{2T/r_0}\right\}\label{BH_collapsing_star_relation2}
\end{eqnarray}
 Substituting the above expression in the equation of the escape angle of a black hole in \eqref{sin alpha_sh}, we obtain
 \begin{equation}
    \sin{\alpha_{sh}^{cs}} = \dfrac{R_s}{R_O} \frac{e^{T/r_0}}{\left|E\right|} \sqrt{1-\displaystyle\sum_{n\in\mathbb{Z}} a_n \dfrac{r_0^n}{R_O^n} - \dfrac{2R_O}{r_0}} \label{escape_angle_cs}
\end{equation}

where the superscript $cs$ stands for the shadow of the collapsing star. Using the expressions of $\Dot{T}$ in \eqref{dTdtau} and $\Dot{R}$ in \eqref{dRdtau} one can in principle obtain $\left|E\right|$  from the energy of the infalling observer when the process of collapse starts, let's say at radius $R_m = R_i$. Substituting this in Eq. \eqref{escape_angle_cs} we get the final expression of the escape angle of the collapsing star as seen by a static observer in $(T,R)$ coordinate system.

\subsection{Analysis in the original Eddington--Finkelstein-like Coordinates}

In this section, our aim is to find the outcomes in the Eddington--Finkelstein coordinates. We perform a coordinate transformation back to the original coordinate system $(u,r,\theta,\phi)$, which will eventually enable us to examine the characteristics of the collapsing star shadow from the viewpoint of an observer sitting on a line with constant ($r,\theta,\phi$). Using the inverse coordinate transformation and 
considering the radius of the star at time $u(T)$ is $r_s = R_s e^{T/r_0}$ we obtain the escape angle of the collapsing star as
\begin{equation}
    \sin{\alpha_{sh}^{cs}} = \dfrac{r_s u}{r_0 r_O \left|E\right|} \sqrt{1-\displaystyle\sum_{n\in\mathbb{Z}} \dfrac{a_n u^n}{r_O^n} - \dfrac{2r_O}{u}} \label{1escape_angle_cs1}
\end{equation}

 where the photon orbit is situated at $r_m$ and the radial position of the observer is at $r_O$. Rescaling $u/r_0$, $r_s/r_0$ and $r_O/r_0$ ({\textit{i.e.}} setting $r_0$ to unity) we get the angular radius of the shadow of a collapsing star from the following relation
\begin{equation}
    \sin{\alpha_{sh}^{cs}} = \dfrac{r_s}{r_O}\dfrac{u}{\left|E\right|} \sqrt{1-\displaystyle\sum_{n\in\mathbb{Z}} \dfrac{a_n u^n}{r_O^n} - \dfrac{2r_O}{u}}. \label{1escape_angle_cs_final}
\end{equation}

The only issue that can arise regarding the evaluation of $r_s$ at any particular instant of time, can be managed by studying the collapse of the star (or alternatively by studying a radial timelike inward geodesic in this spacetime) with appropriate physical conditions (which has been studied by Schneider and Perlick \cite{shadow_dark_star} in $4$-dimensions for the Schwarzschild metric at the exterior).

\section{Simulating a collapsing star with a mass function asymptotically approaching Schwarzschild--Tangherlini Black Hole Mass}

The profile of a collapsing dark star surface for an evaporating black hole is studied in this section following the method presented in \cite{expected_and_unexpected_traits}.  We have already emphasized in Section \ref{SectionIV} that a radially infalling timelike geodesic can mimic the collapse of the star surface. It will be sufficient to study the ingoing geodesics of a massive free particle to visualize 
the collapse. The Lagrangian for a radially infalling massive particle in retarded Eddington--Finkelstein like coordinates is given by
\begin{equation}
    -\left(1-\dfrac{2m_D(u)}{r^{D-3}}\right)\Dot{u}^2 - 2\Dot{u}\Dot{r} = -1 \label{Lagrangian_mass_expel_general1} 
\end{equation}

  Here we choose the Misner--Sharp mass function dependent only on $u$. The equations of motion reduce to the following form with the above choice.
  
\begin{eqnarray}
    \Ddot{r} &+& \left(1-\dfrac{2m_D}{r^{D-3}}\right)\Ddot{u} - \dfrac{1}{r^{D-3}} ~ \dfrac{dm_D}{du} ~ \Dot{u}^2 + \dfrac{2(D-3)m_D}{r^{D-2}} ~ \Dot{u}\Dot{r} = 0 \label{(a)_general} \\
    \Ddot{u} &=& (D-3)\dfrac{m_D}{r^{D-2}} ~ \Dot{u}^2 \label{(b)_general}
\end{eqnarray}

Substituting in Eq. \eqref{(a)_general} the expression of $\Ddot{u}$ from Eq. \eqref{(b)_general} and $\left(1-\dfrac{2m_D}{r^{D-3}}\right)\Dot{u}^2$ from Eq. \eqref{Lagrangian_mass_expel_general1} we get a simplified version of the same equation


\begin{equation}
    \Ddot{r} + (D-3)\dfrac{m_D}{r^{D-2}} - \dfrac{1}{r^{D-3}} ~ \dfrac{dm_D}{du} ~ \Dot{u}^2 = 0 \label{rddot_general}
\end{equation}

We choose an ansatz for the mass function to proceed further with the analysis \cite{Sumanta_dynamic_spacetime_shadow}. 
The choice has been made in such a way that the star collapses to form a spherically symmetric, static black hole (which has been
considered to be Schwarzschild-Tangherlini black hole) at future infinity. The mass function approaches a constant value $m_D^{bh}$ at a  sufficiently large time, {\textit{i.e.}} $m_D(u\to-\infty) \to m_D^{bh} = G_D M_D^{bh}$, where $M_D^{bh}$ is interpreted as the (resulting) black hole mass in $D$-dimensions. A detailed 
discussion on this condition can be found in \cite{Sumanta_dynamic_spacetime_shadow,Coudray_2021}. To proceed further with the numerical studies we have to choose a particular value of the mass $M_D^{bh}$. Let us choose it to be \emph{unity} for the sake of simplicity. However, it is worthwhile to mention that the analysis will remain unaffected for any other value of $M_D^{bh}$. Hence

\begin{equation}
    m_D(u) = \dfrac{1}{2}m_D^{bh} \left(2 + \sech{u}\right) = \dfrac{1}{2}G_D \left(2 + \sech{u}\right) \label{mass_function_case1}
\end{equation}

The evolution of the mass function and the expel profile (represented by the derivative of the mass function) over the course of time for a collapsing star obeying Vaidya metric at the exterior has been shown in the Fig. (\ref{fig1}) below. The mass function is peaked at $u = 0$ and gradually decreases as one probes deep into negative $u$ direction. 

\begin{figure}[htb]
\centering
\includegraphics[width=0.5\textwidth]{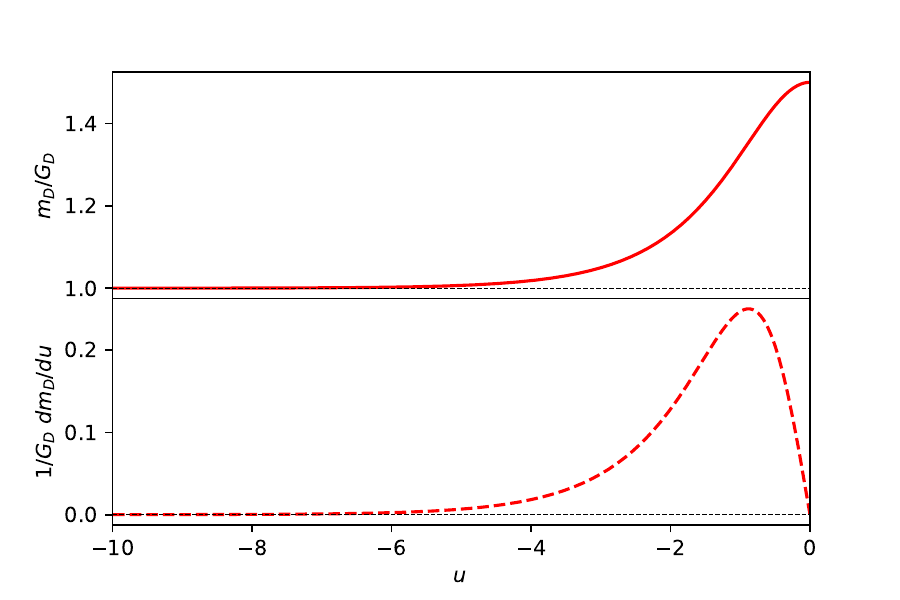}
\caption{Evolution of Miser--Sharp mass parameter, $m_D(u)$ and its time derivative}
\label{fig1}
\end{figure}

Now we find the solution of Eqs. \eqref{(b)_general} and \eqref{rddot_general} subject to the initial conditions -- (i) at $s = u = 0$, $r = r_i$ ($r_i$ is the initial radius of the star) and (ii) $\Dot{u} = \Dot{r} = 0$. The solution has been shown below for $D = 4, 5 ~\text{and}~ 6$-dimensions. The typical size of extra dimension has been chosen as $\mathcal{R}$. Due to lack of our knowledge about the exact size of the higher dimension,  we have plotted  $r(s)/\mathcal{R}$ versus $s/\mathcal{R}$ instead of plotting only $s$ or $r(s)$.  

\begin{figure}[htb]
\centering
\includegraphics[width=0.55\textwidth]{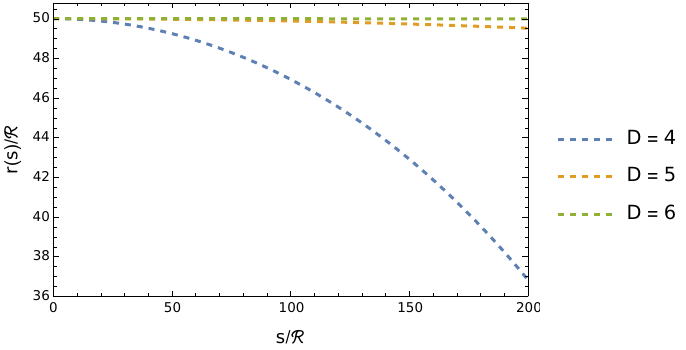}
\caption{Radius {\textit{vs.}} proper time diagram of a collapsing star. The star starts collapsing from outgoing Eddington–Finkelstein time $u = 0$ with radius $r_i = 50$. }
\label{fig:star_surface_collapse_profile}
\end{figure}

It is apparent from the graph that the rate of collapse is faster in four dimensions compared to higher dimensions. The figure (\ref{fig:star_surface_collapse_profile}) shows a star with same initial radius $\left(r_i = 50\right)$ will collapse more rapidly in four-dimensions than in the higher dimension(s). A more massive star in higher dimension will exhibit same rate of collapse compared to a less heavy star in four dimensions. This is, undoubtedly, a very interesting as well as important feature. We now explore the shadow and photon sphere cast by this collapsing star. For this we refer to the general Vaidya metric \eqref{metric}:
\begin{equation}
    ds_D^2 = -f_D(u,r)du^2 - 2 dudr + r^2 d\Omega_{D-2}^2 \label{metric_general1}
\end{equation}

where $f_D(u,r) \equiv 1-\dfrac{2m_D(u)}{r^{D-3}}$. 
The position of the photon sphere and the corresponding shadow in $4D$ have been obtained in \cite{Sumanta_dynamic_spacetime_shadow}. We generalize the results in $D$-dimensions and extend it to the case of a collapsing star possessing the same metric. Consider  $\alpha$ and $\beta$ as the celestial coordinates that define a two-dimensional celestial plane perpendicular to the observer's line of sight and located at spatial infinity \cite{CelestialPlane}. The resulting shadow configuration turns out to be
\begin{equation}
    \alpha(u)^2 + \beta(u)^2 = \dfrac{r_m^2}{f(u,r_m)} \left[1-\left\{\dfrac{\frac{dr_m}{du}}{f(u,r_m) + \frac{dr_m}{du}}\right\}^2\right] \label{shadow_boundary_general1}
\end{equation}
The photon sphere at $r_m \equiv r_m(u)$ in this spacetime is obtained by solving the following differential equation \cite[Eq. 2.9]{Sumanta_dynamic_spacetime_shadow}
\begin{eqnarray}
    \dfrac{d^2 r_m}{du^2} - \left\{\dfrac{3}{r_m}f(u,r_m) - \dfrac{3}{2} \left.\dfrac{\partial f}{\partial r}\right|_{u,r_m}\right\} \dfrac{dr_m}{du} - \dfrac{2}{r_m} \left(\dfrac{dr_m}{du}\right)^2& \nonumber \\
    + \dfrac{1}{2}\left\{f(u,r_m) \left.\dfrac{\partial f}{\partial r}\right|_{u,r_m} + \left.\dfrac{\partial f}{\partial u}\right|_{u,r_m}\right\} - \dfrac{1}{r_m} f(u,r_m)^2& = 0 \label{photon_sphere_general1}
\end{eqnarray}

Substituting $f(u, r)$ for $D$ dimensions and doing some manipulations we get the following form of the above equation 
\begin{eqnarray}
    \dfrac{d^2r_m}{du^2} &-& 3\left\{\dfrac{1}{r_m} - (D-1) \dfrac{m_D(u)}{r_m^{D-2}}\right\} \dfrac{dr_m}{du} - \dfrac{2}{r_m}\left(\dfrac{dr_m}{du}\right)^2 - \dfrac{1}{r_m} \nonumber \\
    &-& 2(D-1)\dfrac{m_D(u)^2}{r_m^{2D-5}} + (D+1)\dfrac{m_D(u)}{r_m^{D-2}} - \dfrac{1}{r_m^{D-3}} \dfrac{dm_D}{du}
    = 0 \label{photon_sphere_general(1)1}
\end{eqnarray}

The time evolution of the photon sphere of a Vaidya spacetime, possessing mass function \eqref{mass_function_case1}, has been studied by solving Eq. \eqref{photon_sphere_general(1)1}. 
We impose the asymptotic condition -- that after a sufficiently large time, the star will collapse to form a Schwarzschild--Tangherlini black hole to solve the above equation. The mass function is chosen accordingly such that it follows the above condition. 
\begin{eqnarray}
        \displaystyle\lim_{u\to-\infty} r_m(u) &=& \left[\dfrac{4\Gamma{\left(\frac{D-1}{2}\right)}G_D}{\pi^{\frac{D-3}{2}}} \left(\dfrac{D-1}{D-2}\right)\right]^{\frac{1}{D-3}} \\
        \text{and} ~~ \displaystyle\lim_{u\to-\infty} \dfrac{dr_m(u)}{du} &=& 0
\end{eqnarray}

\begin{figure}[H]
\centering
\includegraphics[width=7 cm, height= 4 cm]{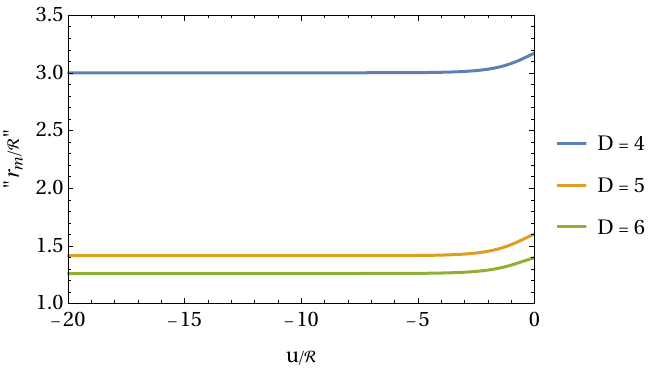}
\caption{Evolution of the photon sphere over the course of time}
\label{fig:photon_sphere}
\end{figure}

From the above conditions we find that a Schwarzschild--Tangherlini black hole has its photon sphere at a position given by the first condition \cite{Sch_Tangh_photon_sphere_derivation},\cite{Schwarzschild_Tangherlini_shadow} and the second condition indicates that the position is not changing over the course of time. We have shown time variation of the position of photon sphere of a Vaidya black hole with mass function \eqref{mass_function_case1} in Figure (\ref{fig:photon_sphere}) where $\mathcal{R}$ represents the size of each extra dimension. As evident from the graph, the photon sphere will shrink more slowly as we go to higher dimension(s). To combat with our lack of knowledge about the exact size of the extra dimension, we have plotted $u/\mathcal{R}$ in the $x$-axis and $r_m/\mathcal{R}$ in the $y$-axis instead of plotting only $u$ or $r_m$.

The relative time evolution of the photon sphere can be visualized more clearly from the evolution of the corresponding collapsing star shadow with $u$. Here we have substituted numerical values of $r_m$ and its first derivatives in Eq. \eqref{shadow_boundary_general1} to plot the shadow at nine different snapshots of time.
\begin{figure}[H]
\centering
\subfloat{\includegraphics[width=.4\linewidth]{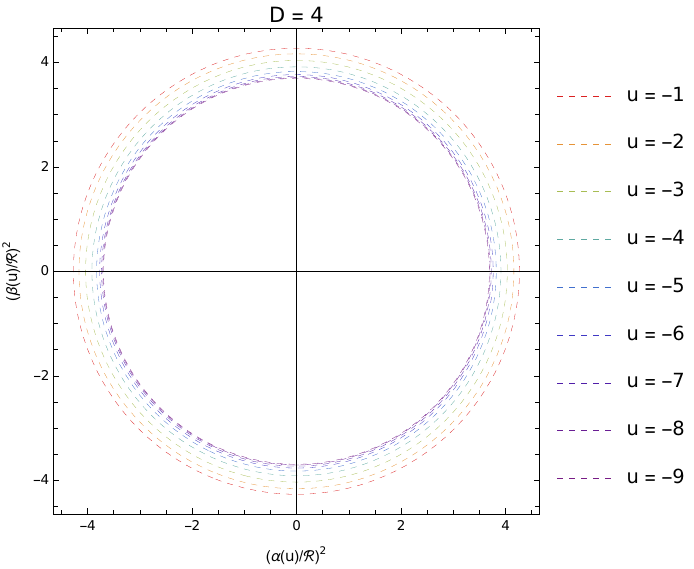}} \par
\subfloat{\includegraphics[width=.4\linewidth]{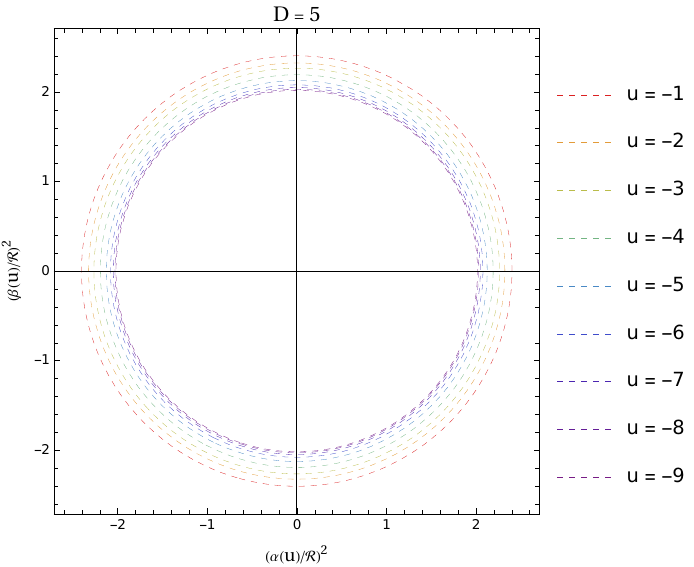}} \hspace{1.8mm}
\subfloat{\includegraphics[width=.4\linewidth]{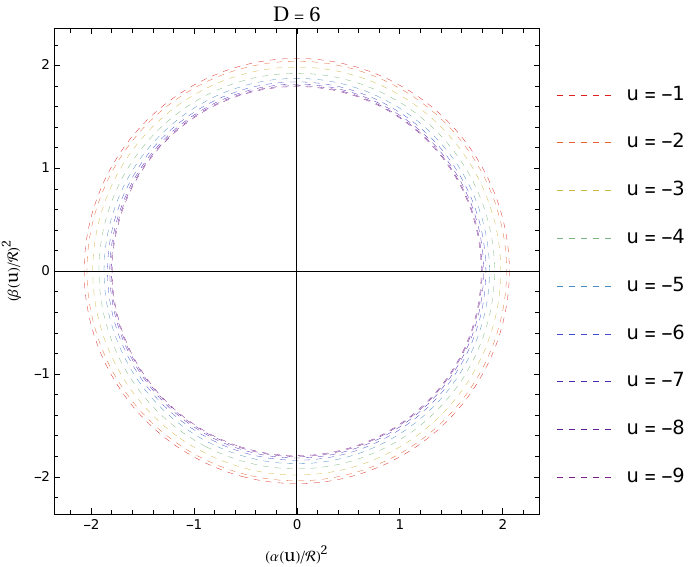}}
\caption{Profile of collapsing star shadow for the same set of physical conditions, but in different dimensions}
\label{fig:shadow_profile}
\end{figure}

In the plots in Fig. (\ref{fig:shadow_profile}) each circle represents the boundary of the shadow at a particular Eddington-Finkelstein time. This can be seen from the figures that in a fixed time interval, the rate of shrinking of the shadow boundary appears to be slower with the increment in the dimensionality of the spacetime. Therefore, in spite of the \emph{similar} physical conditions ({\textit{i.e.}} the boundary conditions) in all the cases, the collapsing star shadow shrinks more faster in $4$-dimensions than that in the higher dimensions. This is quite expected and obvious from our previous analyses in Fig. \ref{fig:star_surface_collapse_profile} where it has been shown that the rate of collapse of the star surface is faster in four dimensions compared to the higher dimensions and therefore the rate of the collapse of shadow profile will behave accordingly.

\section{Signature of Extra Spatial Dimensions} \label{extra_dim_sign}


Let us critically examine the analytical results obtained in the preceding sections. It is evident from Eq. \eqref{photon_sphere_general(1)1} that the shadow of a Vaidya collapsing star (as well as of a Vaidya black hole) strongly depends on the mass and the dimensionality of spacetime. If the position of the photon sphere is kept same (as calculated from the observational data) and dimension $(D)$ is changed, the mass parameter $m_D$ has to be modified accordingly. Therefore, the mass parameter turns out to be an important parameter characterizing the \emph{same} shadow in different dimensions and we can use this property to extract the dimensionality of the spacetime.
To justify our claim we will choose the mass function as in \eqref{m_D1} to explore what are the different values of the mass parameter are responsible for casting the same shadow in $D=4$ and $D>4$. The possibility of existence of such a shadow degeneracy and \emph{isospectral} spacetimes have been discussed in \cite{different_bh_cast_same_shadow} but there all the \emph{isospectral} spacetimes are in four dimensions and the approach is slightly different. \\


 We now ask the obvious question -- whether the extra dimensions can be observationally detected from the isospectrality of spacetime? Or in other words, if different masses cast the same shadow in different dimensions and by some means we get to know which of those theoretically calculated masses is the actual mass of the star -- can we detect the extra dimensions? We present a proposal to give a positive answer to the above question. 
To demonstrate this, we find the relationship between the mass of the star in $D$-dimensional spacetime and the mass in four dimension,  that will cast the identical shadow. Note that we are proposing just a model. So, instead of going into the complications of establishing the relationship between a general mass parameter and the dimensionality of spacetime, we only talk about the analytical form of the mass parameter, that has been considered earlier \eqref{m_D1}.

\begin{equation}
    \dfrac{m_D}{m_4} = \dfrac{r_O^{D-3} \displaystyle\sum a_n \left(u/r_O\right)^n}{r_O \displaystyle\sum a_n \left(u/r_O\right)^n} = r_O^{D-4}
\end{equation}

Our aim is to get {\it an order of magnitude estimation} of the upper bound of the ratio $M_D/M_4$. It is judicious to adopt a sufficiently generalized approach that includes all the additional dimensions with \emph{almost} of the same magnitude, particularly in the context of our specific mass function (Eq. \eqref{m_D1}). Therefore,

\begin{equation}
    M_D \sim M_4 \left(\dfrac{r_O}{\mathcal{R}}\right)^{D-4}
    \label{fig:mass_relation}
\end{equation}
where Newton's constant in $D$ and four dimensions are related by the relation $G_D \sim G_4 \mathcal{R}^{D-4}$ for $\mathcal{R}$ being the typical size of each of the extra dimensions. Using the current bound from LHC we present a crude estimate on the mass $M_{D}$. Substituting the highest achievable energy scale of LHC, i.e. $\sim 13.6$ TeV we derive an upper bound on the size $ \mathcal{R}$ as 
\begin{equation}
    \mathcal{R} < \dfrac{hc}{13.6 ~ \text{TeV}} = 9.12 \times 10^{-23} ~ \text{km}
\end{equation}
From Eq. \ref{fig:mass_relation} we find a lower bound on mass in $D > 4$ dimensions
\begin{equation}
    M_D > \left(3.56 \times 10^5 ~ r_O\right)^{D-4} M_4  
    \label{MD_M4_kpc}
\end{equation}

Note that the typical size of a standard galaxy lies around a few kpc. That is why we have chosen $r_O$ in kpc. Now to determine $M_D$ or $M_4$ which one is actually the mass of that collapsing star we have to rely on astrophysical methods that are generally used to determine the properties of various dark objects in the sky.  \\

It is worth noting that the particular shadow given by Eq. \eqref{1escape_angle_cs_final} can also be cast by some other completely different spacetime. The degeneracy is a serious issue here. We plan to explore this in detail in a future study. The focus of present work was investigation of the $4$-dimensional isospectral counterpart of the higher dimensional spacetime \eqref{metric} and presenting  the method of how the higher dimensional feature of the spacetime can be extracted. The treatment is quite general and can be used in specific case studies. However, different geometry and size of higher dimensions will modify the shadow calculation and will have different outcome. 


\section{Concluding Remarks}

In this work, higher dimensional Vaidya metric has been considered to model a collapsing dark star. 
The {\textit{analytical}} part of the work is restricted to a special type of mass function for which we have calculated  the escape angle and the position of the photon sphere for black hole and collapsing dark star. We have also explored {\textit{numerically}} shadow of the black hole and the collapsing star for an explicit choice of mass function, which is  extendable for any mass function under certain physical conditions. This has been elaborated in the text.
We have further shown how this shadow can be used to study the possible signatures and distinctive features of higher dimensions. 
The analysis is done for a spherically symmetric collapse, however, this method can also be applied for a (slowly) rotating dark star. The major issues along this direction are two fold - (i) lack of a suitable metric around a collapsing rotating star and (ii) of course the mathematical complications involved. The second one can still be handled somehow, but the first one is a major issue. \\

The procedure we discussed here can also be  applied to search the higher spatial dimensions from a black hole shadow \cite{Higher_dim_effect_on_BH_shadow,Schwarzschild_Tangherlini_shadow,Myers_Perry_5D,Higher_dim_BH_shadow1,Higher_dim_BH_shadow2,Higher_dim_BH_shadow3}. However, in the case of a black hole shadow the difference between the masses in different dimensions are generally much smaller and the effects are observationally less prominent than that in our case. The mass of the star has to be much different in four-dimensions than that in the higher dimensions for the formation of same shadow by the collapsing dark star. This makes it much easier from the astrophysical point of view to identify the correct one using other physical parameters as  mentioned in Section \ref{extra_dim_sign}. Overall the process of studying the shadow of a collapsing dark star is very much complex compared to a blackhole shadow. As the Vaidya metric is very general and our work is based on this geometry, the methodology presented here is widely applicable.
Avenues for future research include applying this method to a more realistic and phenomenologically driven mass function, such as the (relativistic) Bondi–Hoyle–Lyttleton mass function. Subsequently, comparing the obtained results with the EHT data could provide conclusive insights into the dimensionality of the spacetime.\\

\medskip
\bibliography{main}
\end{document}